\documentclass[preprint]{aastex}

\slugcomment{Submitted to A.J.}

\shorttitle{On the Formation of Old globular clusters \& their Sites of 
Formation}
\shortauthors{Burgarella et al.}

\begin{document}


\title{Globular cluster systems {\rm II}: On the formation of old globular 
clusters and their sites of formation}


\author{Denis Burgarella}
\affil{Laboratoire d'Astrophysique de Marseille, BP 8, 13376 
Marseille Cedex 12, France}
\email{denis.burgarella@astrsp-mrs.fr}

\author{Markus Kissler-Patig}
\affil{European Southern Observatory, Garching bei 
M\"unchen, 85748, Germany}
\email{mkissler@eso.org}

\and

\author{V\'eronique Buat}
\affil{Laboratoire d'Astrophysique de Marseille, BP 8, 13376 
Marseille Cedex 12, France}
\email{veronique.buat@astrsp-mrs.fr}



\begin{abstract}
We studied the metal-poor globular cluster populations of a large variety
of galaxies and compared their mean metallicity with the properties of the
host galaxies. For this purpose,
we constructed a comprehensive database of old, metal-poor globular cluster
populations, hosted by 47~galaxies spanning about 10~magnitudes in absolute
brightness. The
mean metallicities of the systems are found to be very similar and to lie 
in the $-1.65<$[Fe/H]$\le-1.20$ range (74~\% of the population). Using only 
globular cluster systems with more than 6 objects detected, we find that 
85 \% of the population are within $-1.65<$[Fe/H]$\le-1.20$.  
The relation between the mean metallicity of the metal-poor globular cluster 
systems and the absolute V magnitude of their host galaxies
presents a very low slope which includes zero.  An analysis of the
correlation of the mean metallicity of the populations with other galaxy 
properties (such as velocity dispersion, metallicity, environment
density) also leads to the conclusion that no strong correlation exists. 
The lack of correlation with galaxy properties suggests a formation of
all metal-poor globular clusters in very similar gas fragments.
A weak correlation (to be confirmed) might exist between mean
metallicity of the metal-poor clusters and host galaxy metallicity. This
would imply that at least some fragments in which metal-poor globular
clusters formed were already embedded in the larger dark matter halo of
the final galaxy (as oppose to being independent satellites that were
accreted later).
Our result suggests a homogeneous formation of metal-poor globular
clusters in all galaxies, in typical fragments of masses around 
$10^9$-$10^{10}$ M$_\odot$ with very similar metallicities, compatible with 
hierarchical formation scenarios for galaxies.

We further compare the mean metallicities of the metal-poor globular cluster
populations with the typical metallicities of high redshift objects. If
we add the constraint that globular clusters need a high column density of 
gas to form, Damped Lyman $\alpha$ systems are the most likely sites among the
known high redshift objects for the formation of metal-poor globular cluster 
populations. 
\end{abstract}



%% In the first two sections, notice the use of the natbib \citet
%% and \citep commands to identify citations.  The citations are

%%%%%%%%%%%%%%%%%%%%%%%%%%%%%%%%%%%%%%%%%%%%%%%%%%%%%%%%%%%%%%%%%%%%%%%%%
%%%%%%%%%%%%%%%%%%%%%%%%%%%%%%%%%%%%%%%%%%%%%%%%%%%%%%%%%%%%%%%%%%%%%%%%%

\section{Introduction}

Globular clusters contain the oldest known stellar populations of the Milky Way
and probably of the observed Universe.  Consequently, they hold
cosmologically significant information and are often used as fossil records of
the formation of galaxies. 
At least two globular cluster sub-systems were found to coexist in our Galaxy
\citep{Kinman59,Zinn85,ArmanZinn88,Arman89}: i) a metal-poor, slowly rotating
spherically distributed one in the halo and ii) a metal-rich, rapidly rotating
population concentrated in the disk \citep[or the bulge, e.g.~][]{Minniti95}. 
A similar situation was observed in M~31 \citep{Ashman93,Barmby00}.
Subsequently, globular cluster {\it sub}-populations were also discovered in 
early-type galaxies \citep{Zepf93}. This motivated comparisons between
the globular cluster sub-populations and the host galaxies. Such
studies were pioneered by \citet{Ashman93} and revived by
\citet{Forbes97a}. 
Our work is motivated by the recent increase in data 
identifying metal-poor globular cluster populations, especially in 
early-type galaxies, as well as by recent studies associating the metal-poor
sub-population in early-type galaxies with a extended, halo-like(?) component, 
similar to the ones observed in late-type galaxies 
\citep{Geisler96,Kissler97a,Lee98}. 
This encouraged us to look not only into common properties of these 
sub-systems and to investigate possible correlations between their properties
and those of their host galaxies, but also to put them into a galaxy formation
context and to identify their possible sites of formation.

%%%%%%%%%%%%%%%%%%%%%%%%%%%%%%%%%%%%%%%%%%%%%%%%%%%%%%%%%%%%%%%%%%%%%%%%%
%%%%%%%%%%%%%%%%%%%%%%%%%%%%%%%%%%%%%%%%%%%%%%%%%%%%%%%%%%%%%%%%%%%%%%%%%

\section{The compilation of old globular cluster populations}

\subsection{Focusing on metal-poor globular clusters}

We intend to select the oldest globular clusters around galaxies, with the
minimum pollution from younger globular clusters. 
Even if absolute
ages are not well defined, halo globular clusters in the Milky Way were shown to
be very old systems and at least older than 10 Gyr  
\citep[e.g.][]{Chaboyer98,Jimenez98}.  Globular clusters with [Fe/H] $\le -1.2$ are
essentially located in the halo but, more importantly, they are likely
to be coeval within less than 1 Gyr 
\citep[see][]{Rosenberg99}.  This also seems to be the case in other Local Group
galaxies \citep{Olszewski96,Sarajedini98}.  Furthermore, spectroscopic studies 
of giant
ellipticals showed that the age of their metal-poor globular clusters
appears indistinguishable within the errors (1 to a few Gyr) from that of the 
Milky Way halo clusters \citep{Kissler98a,Cohen98}. 
Assuming that these globular clusters are among the 
first stellar populations formed in the galaxies, we expect them to
reflect the local pre- or proto-galactic conditions and especially the 
abundances more than 10 Gyr ago.
This paper will focus on the metal-poor globular clusters, assumed to be the
oldest ones. 

It was proposed that, in some galaxies, the metal-poor and 
(at least part of) the metal-rich clusters were coeval within a few Gyr
\citep[e.g.~][]{Ortolani95,Feltzing00,Puzia99}. However, given the fact
that this has not been generally demonstrated, and given that a number
of scenarios predict the metal-rich clusters to be younger than the
metal-poor ones, we will not discuss these in this paper.

\subsection{On colors and metallicities and the detection of bimodality}

Our goal is to identify properties that are commonly measured in all
known metal-poor populations. The mean metallicity of the metal-poor
sub-population in a galaxy (noted as [Fe/H]$_{mp}$ throughout the paper) 
is the only such property currently available for a majority of
galaxies. The following analysis will focus on this property, and by
mean metallicity we actually refer to the peak of the metallicity
distribution of the metal-poor globular clusters sub-population in a
galaxy. 

The determination of the mean metallicity of metal-poor populations
is complicated by two problems when derived from broad-band colors, as
for the majority of our sample: {\it i)} a perfect separation of
the metal-poor and metal-rich populations (and thus a determination of
the mean color/metallicity for an unbiased sample of metal-poor clusters)
is not possible from the broad-band color distributions, and {\it ii)} 
the transformation of the broad-band colors into metallicities suffers from 
systematic errors. The uncertainties associated with each point are
discussed below.

In order to determine the mean metallicity of the metal-poor clusters from a 
color distribution of a whole system, the distribution is probed by a
mixture modeling test \citep[KMM, see][]{Ashman94} that, among others,
returns the most likely color peaks of the sub-populations.
Typical errors for the peak determination in $(V-I)$ (the most widely used 
color), induced by the KMM method alone, range from
0.01 to 0.05 magnitudes (sensitively dependent on the number of
data points and the intrinsic width of the distributions) which translates 
into errors in the mean [Fe/H] values
of up to 0.25 dex. These errors are typically added to the statistical errors
present in the photometry, and to potential systematic errors in the
sampling of the metal-poor population.  
A large fraction of our metallicities are
obtained from $(V-I)$ colors derived from WFPC2/HST data. Median color
values for the same system can vary from author to author
by up to 0.06 in $(V-I)$ 
\citep[see for instance NGC 4472 in][]{Neilsen99,Puzia99}.
 Moreover, in order to combine the studies in
different bands, and to combine the results derived from colors and from
spectroscopy, the broad-band colors need to be
transformed into metallicities. The sensitivity of a color to metallicity 
transformation varies by almost a factor of two when going from $(V-I)$
over $(B-I)$ to $(C-T_1)$, 
making a homogeneous compilation difficult.  The different
transformations for a given color into metallicity (often derived from the 
Milky Way clusters) can introduce errors of the order of [Fe/H]$\sim0.3$ dex
depending on the exact method used to derive the relation
\citep[e.g.~the comparison in][]{Kissler98b}.

In summary, taking the quoted mean metallicities directly from the
literature could result in a artificial scatter of up to 0.4 dex in 
[Fe/H]$_{mp}$ in the
extreme cases, given the different analyses of the various authors.
Therefore, in addition to the [Fe/H]$_{mp}$ values, we will use the $(V-I)$
values of the blue peak in the globular cluster color distributions, available 
for a subset of our sample. The latter avoids the possibility of introducing
any error related to a different method of deriving metallicities, or
errors associated with the conversion of colors to metallicities. 

\subsection{The data compilation}

The mean metallicity of the metal-poor clusters, as well as their mean
color (when available) are given in Table~\ref{tbl-1} for all galaxies which 
are known, to date, to host a distinct metal-poor cluster population.
We included the data published by Kundu in
his PhD thesis \citep{Kundu99p} for 9 new galaxies (and additional
data for 8 galaxies already in our list).

New values for 7 galaxies, derived from data presented in Paper {\tt I}
\citep{Gebhardt99}, are also added. The data and
reduction are presented in the original paper.  We selected all galaxies with
clear bimodalities (see Paper {\tt I}), and used the KMM code to derive
the peak color of the blue globular clusters. These values, as well as
the ones from \citet{Kundu99p} were transformed
into metallicities using the ([Fe/H], $(V-I)$ relation given in 
\citet{Kissler98b}: [Fe/H]$=3.27\cdot (V-I)-4.50$.

All other values in Table~\ref{tbl-1} are taken from the original references. 
We also added
to the above sample a number of dwarf galaxies which present a unimodal
metallicity distribution function with an average metallicity below the threshold
defined in Sect.~2.1. We have
assigned global uncertainties to each color~:  $\sigma_{(V-I)}\sim 0.25$,
$\sigma_{(B-I)} \sim 0.20$, $\sigma_{(C-T_1)}\sim 0.15$  
\citep[following][]{Forbes97a} unless different values are given in the
original papers.

The compilation includes galaxies of all types. However spiral galaxies are
under-represented and bright elliptical galaxies dominate the sample.
This observational bias is essentially due to the fact that~: {\it i)} globular
clusters are more easily identified on the smooth background of elliptical
galaxies than in dusty spirals, {\it ii)} bright ellipticals host a larger
number of globular clusters than faint ellipticals or spirals.  

The host galaxy properties to which we compare the globular cluster properties
are compiled in Table~\ref{tbl-2}.  These are taken from the {\tt HYPERCAT} database
\citep{Prugniel96,Golev98},
except for the environment density taken from the Nearby Galaxy Catalog
\citep{Tully88}.  We chose the Mg$_2$ index as a metallicity
indicator rather than, for example, the color of the galaxy, because it is 
widely available and does not require dereddening or transformation to
reflect the metal content of the galaxy. We keep in mind that
Mg$_2$ is essentially measured in the very inner regions of the galaxy,
and does not directly reflect the mean metallicity of the halo.
Nevertheless, it appears to be a
good indicator for the final global metallicity and correlates well with
the velocity dispersion of the galaxy \citep[e.g.][]{Dressler87}.  

The above mentioned velocity dispersion is used as a size
indicator for the galaxies, rather than e.g.~the estimated absolute
magnitude, since the former is distance independent, while the latter is
not and is therefore more difficult to bring onto a homogeneous scale.  
In order to get 
comparable values for all galaxies, we tried to select only seeing-limited
ground-based determinations for the central velocity dispersion,
e.g.~HST/STIS
values being systematically higher due to the higher spatial resolution.

The data described above is used in the next section to investigate 
possible correlations between the mean metallicity of the globular clusters 
and the host galaxy properties.

%%%%%%%%%%%%%%%%%%%%%%%%%%%%%%%%%%%%%%%%%%%%%%%%%%%%%%%%%%%%%%%%%%%%%%%%%
%%%%%%%%%%%%%%%%%%%%%%%%%%%%%%%%%%%%%%%%%%%%%%%%%%%%%%%%%%%%%%%%%%%%%%%%%

\section{Globular cluster mean metallicities and galaxy properties}

\subsection{A ``universal'' mean metallicity for metal-poor globular clusters}

Until the early 90s, the mean metallicities of the
globular cluster systems was thought to correlate with the galaxy
luminosity \citep{vandenBergh75,Brodie91}.
\citet{Ashman93} first investigated the correlation between
the mean metallicity of the metal-poor globular clusters only and the
galaxy luminosity. They based their analysis on data of 4 local dwarf galaxies 
as well as the LMC, the Milky Way and M31. They found that a mean value
of [Fe/H]$\sim -1.6$ dex for all halo globular cluster systems appeared
consistent with the data, and claimed that genuine halo globular cluster
systems have comparable mean metallicities, irrespective of the parent
galaxy luminosity. Further, adding the data of the 4 early-type
galaxies known at that time to have bimodal globular cluster metallicity 
distributions, they speculated that the earlier relations (see above)
were primarily a result of the high fraction of metal-rich globulars in
bright elliptical galaxies.

\citet{Forbes97a} later confirmed this result with a slightly larger sample of
11 galaxies, by looking at the
correlation of mean metallicity with galaxy luminosity for the metal-poor
and metal-rich clusters separately.  For the metal-rich population, they
found a positive correlation at the 3 $\sigma$ level.  For the
metal-poor globular clusters, \citet{Forbes97a} did not
detect any correlation but a random scattering.  Their dataset has a mean of
[Fe/H]$_{mp}$$=-1.16$ with a dispersion of $\sigma=0.28$ (as
determined by us using a maximum-likelihood estimator on their 11 values).

\begin{figure} 
\plottwo{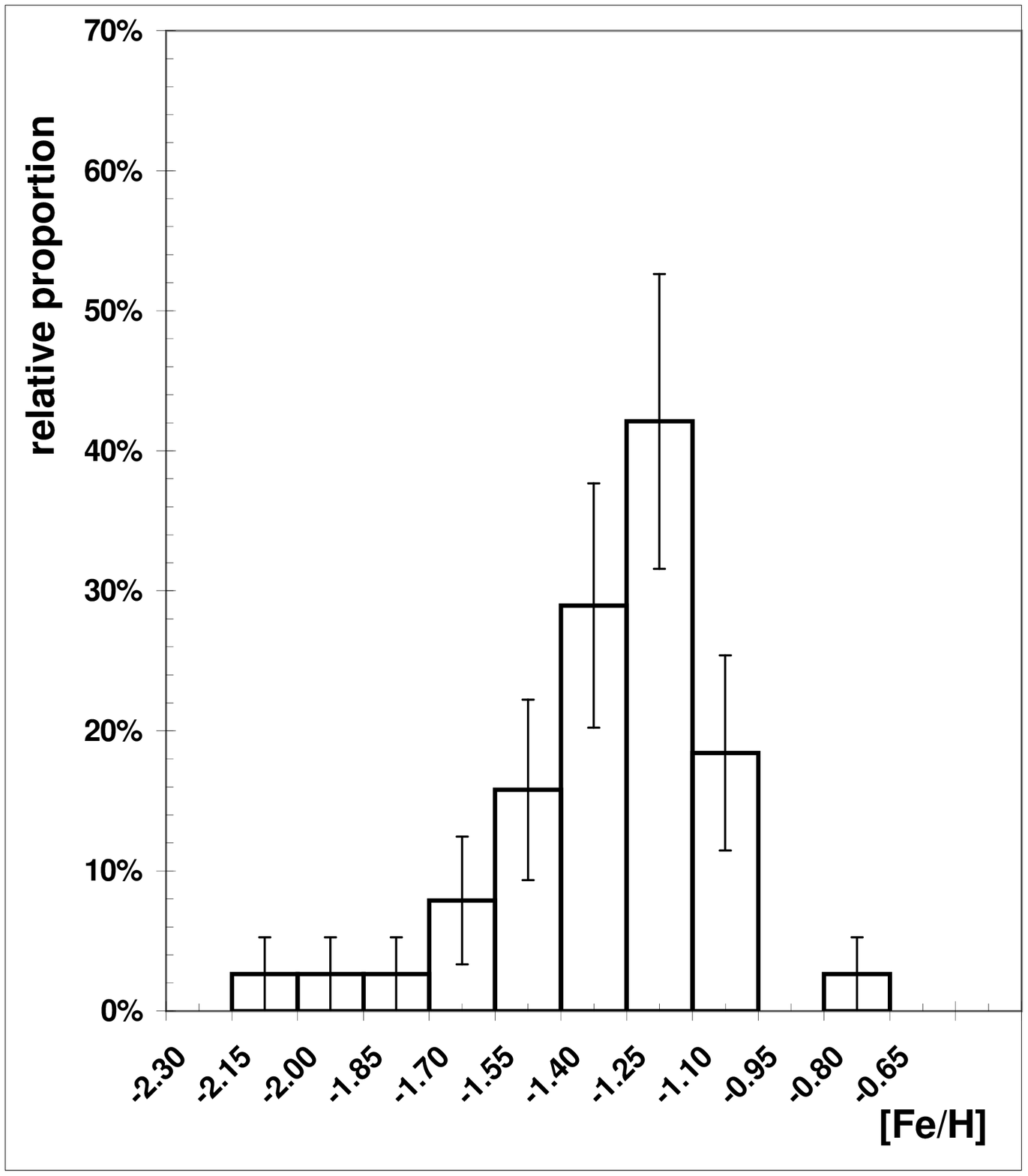}{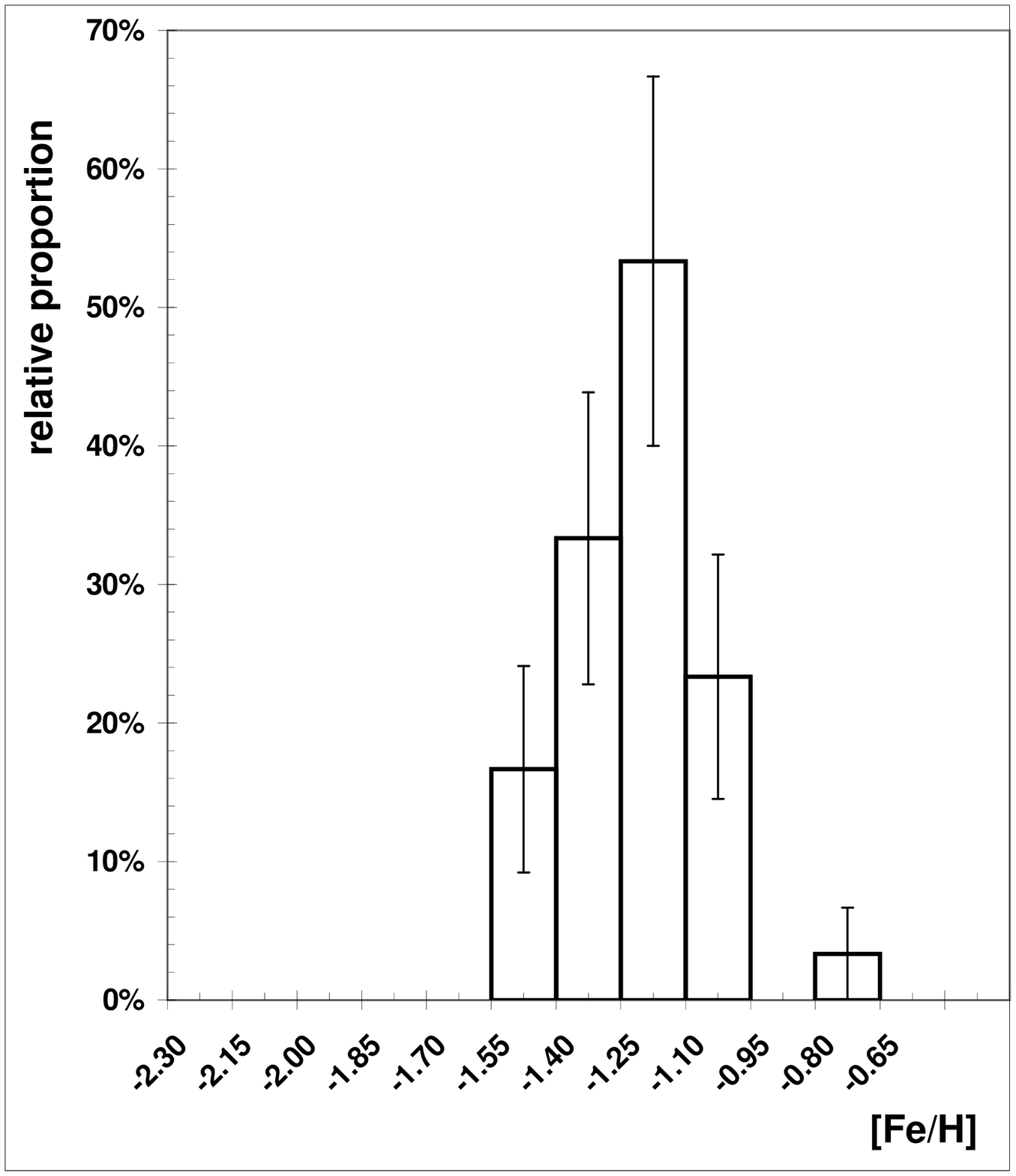}
\caption[]{Distribution of mean metallicities for the globular cluster 
systems. The left-hand panel presents the distribution of the mean
metallicities  of metal-poor clusters for the complete sample and the 
right-hand panel only shows globular cluster systems with $N_{GC}>6$
(see text).\label{fehhisto} } 
\end{figure}

The sample of globular cluster systems presented in this paper is the largest
database to-date, and $>4$ times more numerous than 
\citet{Forbes97a} initial dataset.  The mean of our data lies at
[Fe/H]$_{mp}=-1.45$ with a dispersion of $\sigma=0.15$. Compared to the 
Forbes et al.~dataset, our sample is slightly more metal-poor on average
and exhibits a smaller scatter. The
mean absolute magnitude of the galaxies in our sample is $<M_V> = -20.1$ with a
dispersion of 2.4 mag.  
Fig.~\ref{fehhisto} shows the distribution of mean metallicities,
plotted as a percentage of globular
cluster systems within each bin ($\Delta$[Fe/H]$_{mp}=0.15$).  The
first apparent result is that the mean-metallicities of the metal-poor
globular clusters are not distributed homogeneously over the spanned
range, but rather peak around a characteristic value~: 74\% of the sample
is concentrated around $-1.65<$[Fe/H]$_{mp}\le-1.20$. 
The distribution is asymmetric and more 
extended towards lower metallicity (dominated by the dwarf galaxies),
resembling in shape to the halo field star metallicity distribution.  
When we exclude galaxies with a number of observed blue
globular clusters N$_{GC}\le 6$ (almost exclusively dwarf galaxies) 
from the sample, our statistics are slightly
altered to [Fe/H]$_{mp}\sim -1.37\pm 0.2$ and for the 39 remaining galaxies
$<M_V>=-20.8\pm 1.6$. The new histogram (Fig.~\ref{fehhisto}) appears more 
peaked. Indeed, 85 \% of the globular cluster systems are found within
$-1.65~<$[Fe/H]$_{mp}\le-1.20$ and 67\% within the two central bins 
i.e.~$-1.55<$[Fe/H]$_{mp}\le-1.25$.
Thus, the data suggest an even higher concentration of the metal-poor
globular clusters.

The peaked (roughly Gaussian) distribution of the mean metallicities would be 
expected on the base of the Central Limit Theorem if all metal-poor 
globular clusters could be associated with a single sample
(i.e.~be considered to have a similar origin). The fact that the distribution
looks indeed peaked supports the latter hypothesis.

\begin{figure} 
\plotone{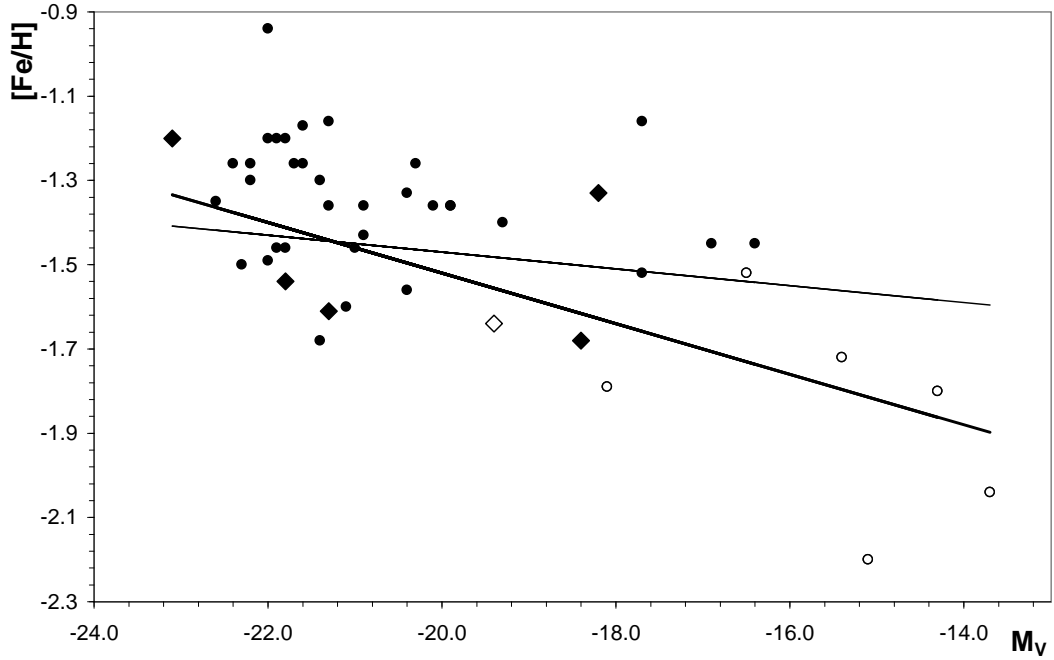}
\caption[]{Mean metallicity of the old, metal-poor globular cluster systems
plotted against the absolute magnitude $M_V$ of the parent galaxy.  Circles
represent E/SO galaxies while  diamonds show spirals. Filled symbols
represent globular cluster systems with an observed number of globular
clusters $N_{GC} > 6$, open symbols stand for systems with $N_{GC} \le 6$. 
The two lines represent linear fits to the complete sample and to the 
sub-sample with $N_{GC}>6$ respectively. Note that their slopes are
not significantly different from zero within the errors. \label{fehmv} }
\end{figure}

Figure~\ref{fehmv} shows the metallicities of the globular cluster
systems as a function of the absolute magnitude $M_V$. 
The global trend is a decrease of the metallicity with $M_V$. The
statistical Spearman's rank test seems to confirm this impression and gives 
a probability of 0.0005 that a correlation is not present 
(Spearman's $\rho$ = -0.521). A linear fit gives~:
[Fe/H]$_{mp} = -0.06 (\pm 0.01) M_V - 2.72 (\pm 0.22)$ for 46 values. 
However, removing the globular cluster systems
with N$_{GC}$ $\le$ 6 from the sample (but keeping Kundu's), the same Spearman 
rank test 
gives a much lower probability of 0.0813 that a correlation is not present 
(Spearman's $\rho$ = -0.283 for 39 values).
A linear fit gives [Fe/H]$_{mp} = -0.02 (\pm 0.02) M_V - 1.87 (\pm
0.31)$, with a low slope not significantly different from zero (at 1$\sigma$).

Thus, the mean metallicity of the old, metal-poor globular clusters seems to
correlate with the absolute luminosity of
their host galaxy. However, taking only the galaxies with N$_{GC}$ $\ge$ 6, 
this correlation is no longer statistically significant while we still
have a large range in galaxy luminosity ($-23<M_V<-16$). Thus, our findings
confirm Ashman \$ Bird's results for the metal-poor clusters. 
However, the globular
cluster systems of dwarf galaxies seems to deserve a more complete discussion
and they will be discussed in a future paper.

\subsection{Mean metallicities against Mg$_2$, $\sigma$, and environment
density}

Next, we investigate whether the mean metallicity of the metal-poor globular
clusters correlates with other galaxy properties, like metallicity, size or
environment.  We used the derived mean metallicities, as well as the mean $(V-I)$
values (when available), to avoid possible systematic errors due to the
different transformations from broad-band colors to metallicities.  Note that
when several values for $(V-I)$ were available for a given system, we computed a
simple mean, and averaged it with any other metallicity determination,
if available for that system.

\begin{figure}
\plotone{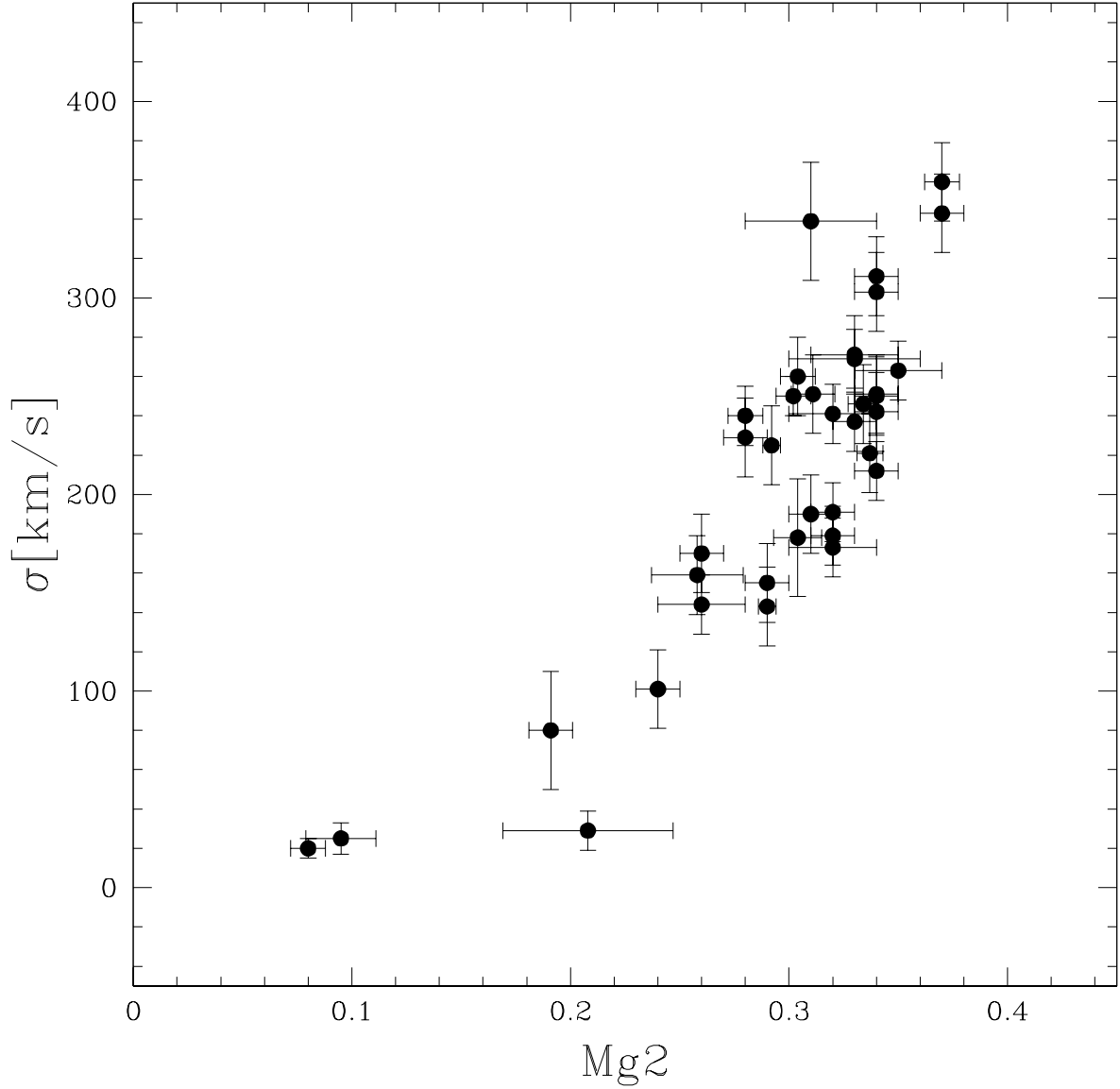}
\caption{
The figure shows the clear correlation between Mg$_2$ and $\sigma$ for
the galaxies of our sample, as expected \citet[e.g.][]{Dressler87}.\label{sigmg}}
\end{figure}

The correlation between [Fe/H]$_{mp}$ and the Mg$_2$ index of the host
galaxy, as well as between  [Fe/H]$_{mp}$ and the velocity dispersion
$\sigma$ of the host galaxy, computed for the full sample, are significant at 
the $>$99\% confidence level (spearman test returning 
student distributed $t$-value$=2.45$ and 2.90 with 35 and 40 degrees of 
freedom, respectively). Mg$_2$ and $\sigma$ being strongly correlated
(see Fig.~\ref{sigmg} for our sample), the 
similarity of the relations is not surprising.
Neglecting, however, the dwarf galaxies (Mg$_2<0.25$, $\sigma <150$
km$\cdot$s$^{-1}$) 
which do not exhibit a separate metal-poor component, reduces the
significance of the correlations to the $<$92\% confidence level
($t=1.45$ and 1.30 for 33 and 35 degrees of freedom, respectively). 
Furthermore, the significance is reduced even further when selecting
the clusters as in Sect.~3.1 (i.e.~only galaxies with more than 6 clusters are
considered). The correlation disappears well below the 90\% confidence level
($t=1.07$ and 0.75 for 32 and 34 degrees of freedom, respectively). 

A linear fit returns
a relation of the form [Fe/H]$_{mp}=1.07(\pm0.44)\cdot{\rm Mg}_2 -1.68
(\pm0.13)$ for the full sample. The slope of this relation is $\sim 10$
times shallower than a direct conversion of Mg$_2$ into [Fe/H]$_{mp}$
\citep[see][]{Kissler98b} indicating only a very weak dependence of
globular cluster mean metallicity on galaxy metallicity, if present at all. 

Again, a possible correlation for dwarf galaxies will be discussed in a future
paper discussing metal-rich globular cluster sub-populations.

We note that absolutely no correlation is detected with environment
density ($t=-0.63$ for 35 degrees of freedom).

\subsection{$(V-I)$ against Mg$_2$, $\sigma$, and environment density}

Similar test as in the above section were preformed for $(V-I)$ instead of
metallicity in order to avoid any potential systematic effects arising
from the conversion of color into metallicity.

\begin{figure}
\plottwo{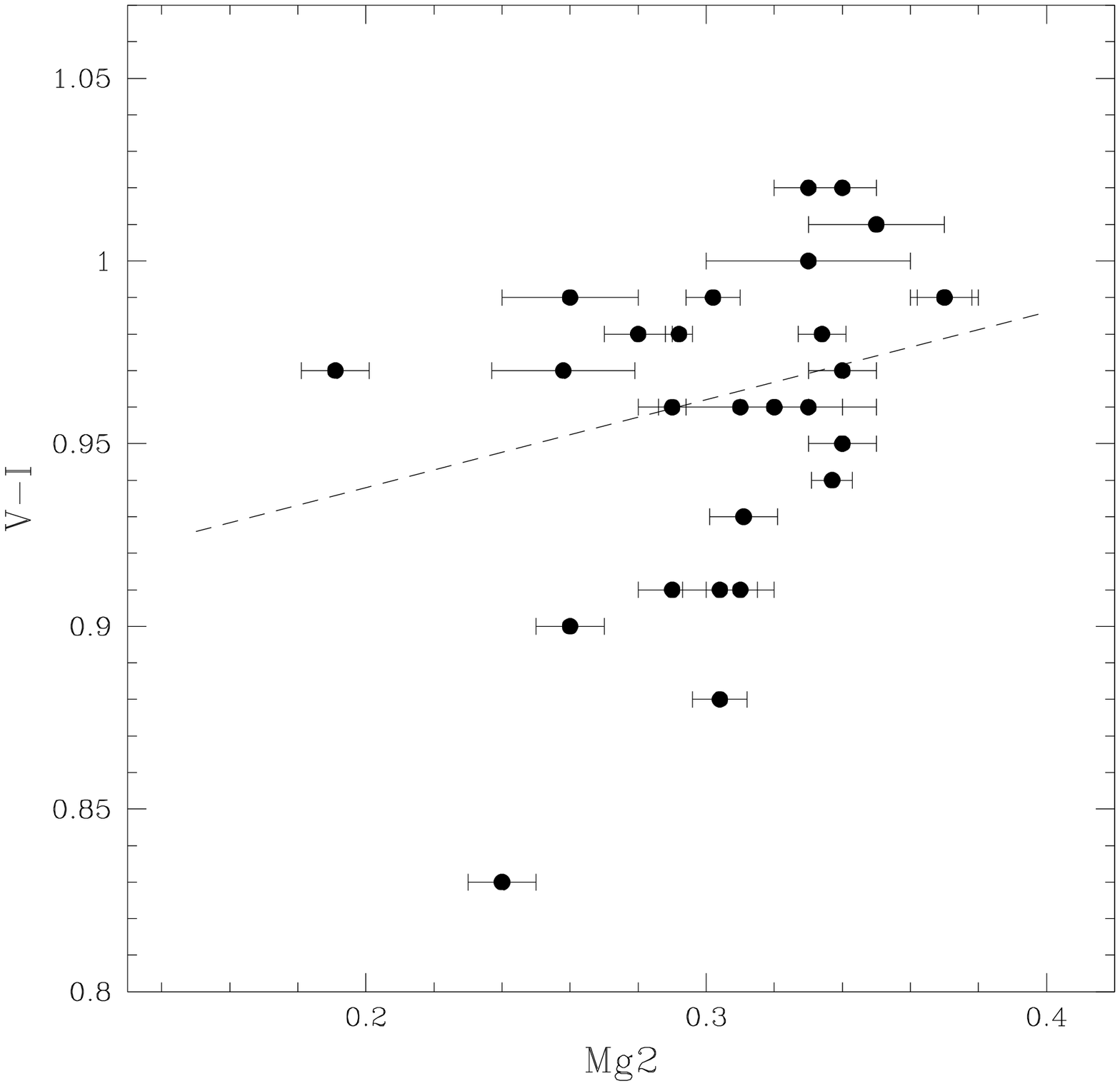}{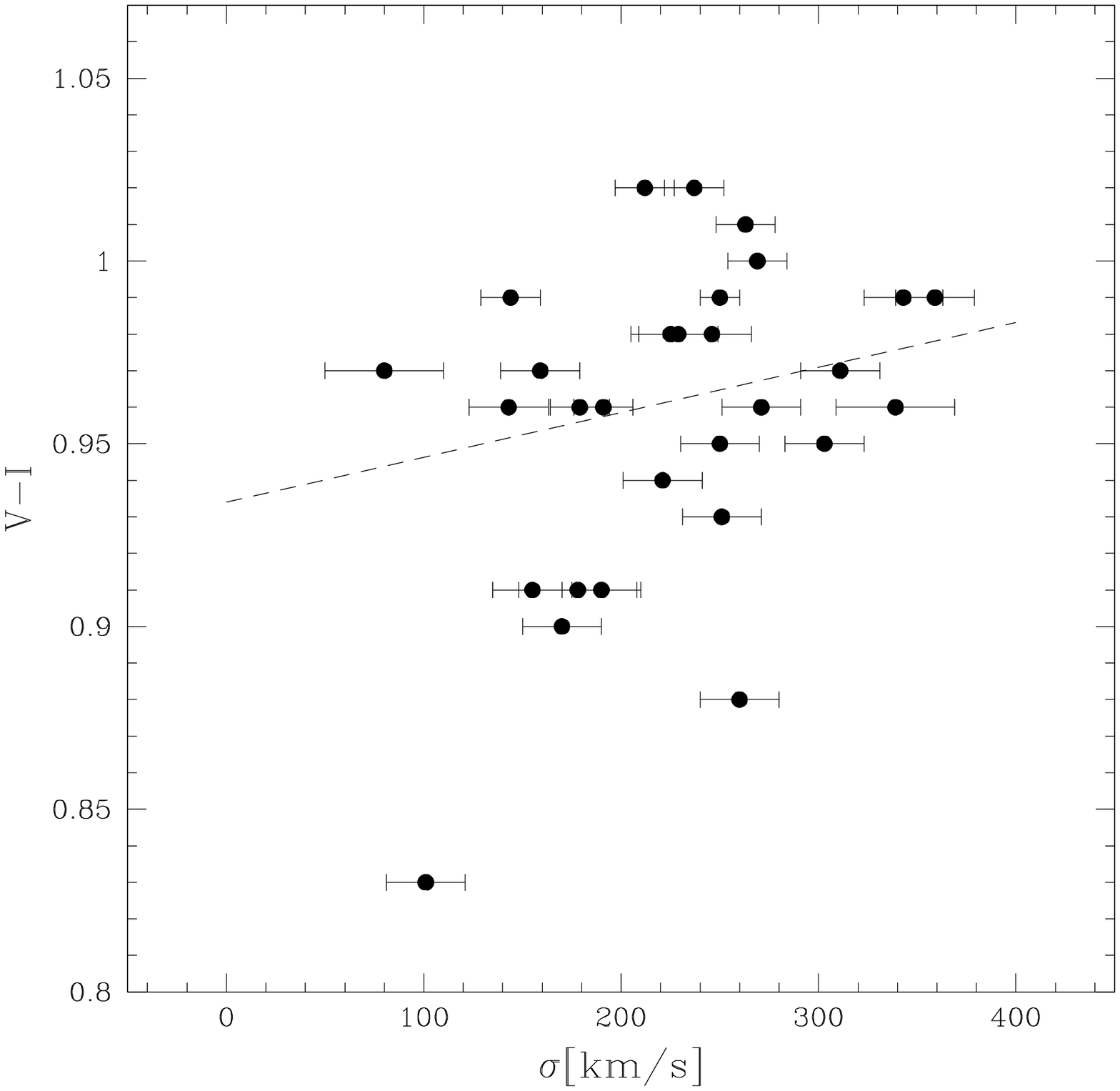}
\caption{
The figure shows the peak $(V-I)$ color of the metal-poor globular cluster
systems plotted against the Mg$_2$ index (left panel) and velocity dispersion
(right panel) of their host galaxies. The dashed lines show a linear
least square fit to the data (excluding the datapoint of NGC 4458). A
weak ($<2\sigma$ confidence) correlation might exist between the plotted
values. \label{vimgsig}
}
\end{figure}

The most significant correlations of the peak color with a galaxy property
are again the correlation between $(V-I)$ and Mg$_2$, and $(V-I)$ and
$\sigma$ for the full sample. The quantities are correlated at the
$\sim$97\% confidence level ($t=1.89$ and 1.80 for 27 degrees of freedom
respectively). These correlations are mostly driven by NGC 4458, the
only galaxy with 6 or less clusters for which a $(V-I)$ peak color is
available. Note, however, that NGC 4458 is
a rather uncertain detection: \citet{Neilsen99} had 17 
clusters to detect a bimodality in the color distribution, of which 6
are associated to the metal-poor peak. Neither \citet{Kundu99p}, nor we 
could detect a bimodality and reproduce this result with 33 clusters detected
in the \citet{Gebhardt99} dataset. Removing this galaxy from the sample,
the significances of the correlations are reduced to the $\sim$94\% and
$\sim$90\% confidence level ($t=1.67$ and 1.39 for 26 degrees of freedom, 
respectively), becoming marginal.
A linear fit returns
a relation of the form $(V-I)=0.24(\pm0.18)\cdot{\rm Mg}_2 -0.89
(\pm0.06)$ for the sample excluding NGC 4458. The slope is
compatible with 0 within 1.3 $\sigma$ errors (a similar result is
obtained for the velocity dispersion). Figure \ref{vimgsig} shows the
relations.

A correlation with environment density is completely absent ($t=0.08$ for 26 
degrees of freedom). 

We conclude that, the mean metallicity of
the metal-poor globular clusters is not significantly related to the
size, metallicity or environment of the host galaxy. 
A weak correlation might exist but remains to be confirmed.

Interestingly, Larsen et al.~(2001 in preparation) find for a very
homogeneous sample of 12 galaxies a similar trend. Tentative
implications are discussed below.

%%%%%%%%%%%%%%%%%%%%%%%%%%%%%%%%%%%%%%%%%%%%%%%%%%%%%%%%%%%%%%%%%%%%%%%%%
%%%%%%%%%%%%%%%%%%%%%%%%%%%%%%%%%%%%%%%%%%%%%%%%%%%%%%%%%%%%%%%%%%%%%%%%%

\section{Some constraints on the formation of halo globular clusters}

From the results of Sect.~3 we retain two important points~:

1) the mean metallicity of our sample of metal-poor globular cluster 
systems is only weakly (if at all) dependent of the host galaxy properties 
(M$_V$, type, 
environment, metallicity). This suggests that the formation of metal-poor 
globular clusters was 
largely uncorrelated with the final host galaxy properties. 
The metal-poor globular clusters (often associated with the halo)
could have formed 
either in the proto-galactic phases of the host galaxies or in the earlier 
phases of the galaxy formation. The old age of the globular clusters 
also supports these latter ideas. 

2) the mean value of our sample of metal-poor globular cluster 
systems seems almost ``universal'' at [Fe/H]$_{mp} \sim -1.4$ with a low
dispersion of the metallicity distribution function of $\sigma_{{\rm
[Fe/H]}_{mp}} \sim 0.3$.
This suggests a ``universal'' mode of formation for the metal-poor
globular clusters, in the sense that their formation sites had very
similar metallicities / properties~(?).

\subsection{Size of the putative fragments}

With this formation hypothesis in mind, we can explore a noteworthy 
consequence on the fragment sizes. \citet{Ashman93} already addressed
this problem comparing globular cluster sub-groups within the M31 halo with 
expectations from cold dark matter models (predicting substructures of the 
order of $10^{-3} M_{\rm halo}$). They found the observations and
predictions in good agreement, with typical mass scales for
substructures within M31 of $2\cdot 10^9$ M$_\odot$.
The sizes can also be derived by combining 
i) the mean metallicity of the metal-poor globular clusters around 
giant galaxies and ii) an assumed relation between the galaxy initial 
luminosity and the average globular cluster metallicity \citep{Cote98}. 
Relating the two, most metal-poor globular clusters
must have formed in objects with luminosities around M$_V\sim -17$, i.e.~masses
 of the order of $10^9$ to $10^{10}$ M$_\odot$ (i.e.~in galaxies larger than 
the remaining dwarfs observed in the Milky Way neighborhood). 
In good agreement with \citet{Ashman93}'s results.

\subsection{Was the formation of the metal-poor globular cluster
independent of the final host galaxy~?}

The current data favor a formation of most metal-poor globular clusters 
in very similar environments/substructures, out of low-metallicity gas 
\citep[as already speculated by][]{Ashman93}.

Whether these substructures were fragments \`a la \citet{Searle78},
i.e.~entities within the dark matter halo of the final host galaxy, or
satellites with similar properties but independent of the dark matter
halo of the final galaxy, is unclear. But it appears secure that they
formed independently of the metal-rich component (bulge) of the host galaxy.

Thus, the properties cannot distinguish yet between a scenario in which
the metal-poor globular clusters formed completely independently of the
final galaxy and were later accreted
\citep[e.g.][]{Richtler94,Cote98,Hilker99}, and a scenario in which they
formed as part of the galaxy during the assembly of the halo
\citep{Searle78}. The latter would be favored if a correlation between the 
mean metallicity of the metal-poor clusters and the metallicity of the host 
galaxy would exist. The former would be favored if no such correlation
would be present.
Both scenarios would not restrict any formation scenario
of the final galaxy (e.g.~major collapse, major merger, ...), since all
galaxy formation scenarios envision similar assemblies of the halos (be
it as a first stage of further collapse or as the halos of progenitors
of a subsequent major merging event).

If existent, the correlation between the mean metallicity
of the metal-poor population and the galaxy metallicity is about 10
times shallower than a one to one relation.
Such a correlation would imply that a fraction of the 
metal-poor globular clusters formed in satellites (i.e. not related
to the dark matter halo of the galaxy) but get accreted later while
the other part formed in fragments (i.e.~already within the 
dark halo of the final galaxy).
For our dataset such a correlation is, however, still uncertain, and the
fraction of metal-poor clusters from both origins in a system that contributes 
to it is also unknown and probably variable with galaxies.

Finally, we do know that, in the Milky Way, some halo globular clusters
formed around other galaxies and were accreted later on. For
example, such an accretion process can be witnessed today in the form of
the Sagittarius dwarf galaxy \citep{Ibata94}.
Although halo globular clusters of the Milky Way form a homogeneous
population from their metallicity, it has been suggested from an analysis of 
their horizontal branch types \citep{Zinn93} that 
it may contain two sub-systems with similar average metallicities. 
This would complicate the interpretation of their origin.
A review of pros and cons can be found in \citet{Ashman98}, \citet{Harris00} and
\citet{Parmentier00}.

%%%%%%%%%%%%%%%%%%%%%%%%%%%%%%%%%%%%%%%%%%%%%%%%%%%%%%%%%%%%%%%%%%%%%%%%%
%%%%%%%%%%%%%%%%%%%%%%%%%%%%%%%%%%%%%%%%%%%%%%%%%%%%%%%%%%%%%%%%%%%%%%%%%

\section{Time and site of formation of the metal-poor globular clusters}

\subsection{DLA systems as the progenitors of metal-poor globular clusters} 

\citet{Pettini99} presented a diagram giving the rough location
of different components of the early universe at z $\sim$ 3 in a N(HI) --
metallicity plane.  Among the objects whose metallicity can be estimated at high
redshift are the Damped Lyman $\alpha$ (DLA) systems, the Lyman Break Galaxies
(LBG) and the Lyman $\alpha$ forest.  DLA systems are neutral gas objects observed at
all redshifts \citep{Pettini99,Prochaska00}.  Their dynamics is consistent with protogalactic clumps or
progenitors of present-day galactic disks \citep{Wolfe95,Katz96,Haehnelt96}.  
One interpretation of DLA systems is that they are gas clouds within
protogalactic halos, that could be associated with Searle-Zinn fragments.
LBGs are star-forming objects similar to our local starbursting galaxies 
\citep{Steidel96}  They appear as objects with compact cores, some with
multiple components, surrounded by diffuse and asymmetric halos 
\citep{Steidel96,Lowenthal97}.  If the metallicity of
the Lyman $\alpha$ forest is definitely too low, DLA systems and LBGs are found to lie
in the same metallicity range as halo globular clusters.  However, recent data
from \citet{Kobulnicky00} show that the mean interstellar
medium metallicity in their LBGs is consistent with the metal-rich globular
cluster population in the Milky Way.  Consequently, since 
metal-poor globular clusters are formed from metal-poor gas,  
DLA systems appear as the best candidates for their site of formation.

Furthermore, globular clusters contain stars and the column density of
their progenitors must be above a threshold 
of N(HI) $\sim$ 10$^{20}$ cm$^{-2}$ for the star formation to occur  
\citep{Kennicutt89}. Simulations carried out by 
\citet{Nakasato00} also suggest that a self-enrichment in a cloud
appears to exclude the formation massive star clusters. However, this point is 
still controversial \citep{Parmentier00}. 

Bringing the above mentioned facts together confirms the allowed location 
of halo globular cluster progenitors in the N(HI) 
vs. [Fe/H] diagram and suggests that we should concentrate, as a working
hypothesis, on DLA systems as the potential sites of formation of the (halo) 
metal-poor globular clusters. 

\subsection {Estimating the metallicity evolution of DLA systems}

\begin{figure} 
\plotone{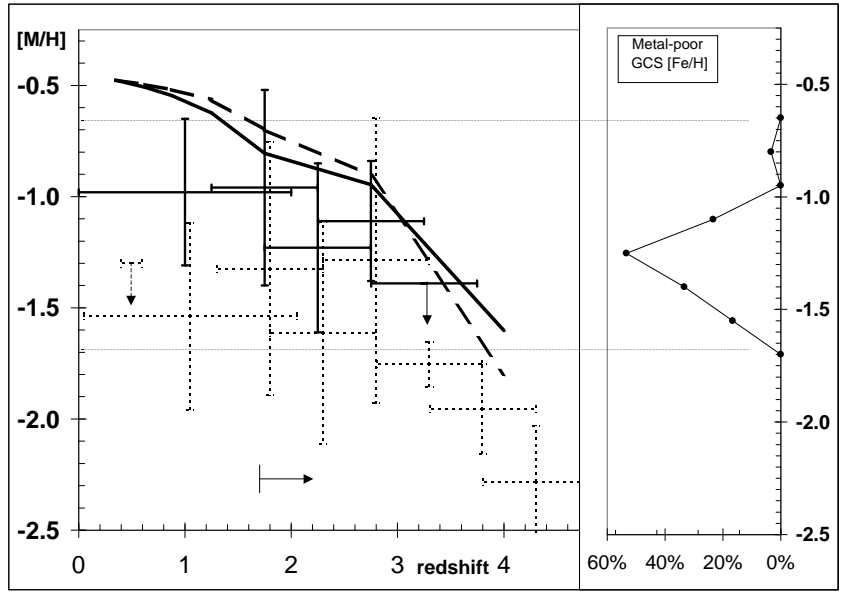}
\caption[]{Comparative variation of the metallicities with the redshift 
(H$_0=50$km$\cdot$s$^{-1}\cdot$Mpc$^{-1}$ and q$_0=0.5$).  Right hand panel:  
globular cluster system metallicity distribution.  Left hand panel : 
the limits of the globular cluster system mean-metallicity
distribution (N$_{GC}>6$) are reported on the left hand plot as 
dashed lines. The uppermost
curves have been deduced from \citet[Fig.~9]{Steidel99}
star formation history (continuous line); the dashed line includes 
\citet{Barger99} FIR data.  [Zn/H] (continuous line crosses) and [Fe/H] 
(dashed crosses) values of DLA systems are taken from Table~\ref{tbl-3}. The age of the 
oldest Galactic globular clusters is assumed to be $>10$ Gyr, and is 
reported as an horizontal right-bound arrow.
\citet{Pettini97} noted that assuming q$_0=0.01$ would shift 
the [M/H] by a factor of 2. \label{gcsdlaz}} 
\end{figure}

With the previous hypothesis that DLA systems can be connected to 
the progenitors of the metal-poor globular clusters, we take advantage of 
the available database of DLA observations, normalized
to \citet{Anders89} solar abundances : Log~[Fe/H]$_\odot =
-4.49$ and Log~[Zn/H]$_\odot=-7.35$, to study the chemical evolution of
these objects with redshift.  However, [Fe/H] may not be a reliable
estimate of the metallicity of DLA systems, since some Fe may be locked up in dust
and result in biased [Fe/H] measurements.  \citet{Pettini97}
showed that [Zn/H] is a more reliable estimator because it essentially
measures the metallicity independently of dust depletion. The [Zn/H]
and [Fe/H] variations as a function of the redshift (Table~\ref{tbl-3}) 
are plotted in Fig.~\ref{gcsdlaz}.
We use the column-density weighted abundances~:  $ \rm [<M/H_{DLA}>] = log
<(M/H)_{DLA}> - ~log (M/H)_\odot $ where $\rm <(M/H)_{DLA}>$ (M = Fe or M =
Zn) and the associated standard deviations as defined in \citet{Pettini97}.

The average values $<[Fe/H]_{DLA}>=-1.53 \pm 0.40$ and
$<[Zn/H]_{DLA}>=-1.13 \pm 0.38$ for the two complete samples, give
$<[Zn/Fe]_{DLA}>=0.40\pm 0.55$ over the whole sample.  We do not
normalize the [Fe/H] to the [Zn/H] values since it remains unclear
whether $<[Zn/Fe]_{DLA}>$ is constant with redshift. 
A similar trend of decreasing metallicity with redshift
appears both for [Fe/H]$_{DLA}$ and [Zn/H]$_{DLA}$.  The possibility that
part of this trend is caused by a fraction of high redshifts DLA systems missed
because of dust is investigated and ruled out by \citet{Pei95} 
\citep[but see also][]{Prochaska00}.  The steeper slope at z $>$ 2.8
has been interpreted as the fast formation of metals after the dark age
\citep{Pettini97,Lu96}.

In order to verify whether the above trend of decreasing metallicity with 
increasing redshift is supported by other evidence, we further compute
the evolution of the 
metallicity of the universe with the redshift. We use the star formation
evolution of \citep{Steidel99} and also consider an
alternative scenario between z = 1 and 3 from sub-mm data 
\citep{Barger99}. The second scenario implies a higher metal production
at z $<$ 3 and a lower metal production at z $>$ 3.  The expected
metallicity is computed following \citet{Pettini97} except
that the values are normalized to give a global metallicity in the
present-day Universe Z = 1/3 Z$_\odot$ \citep{Renzini99}.  As
noticed by \citep{Pettini97}, the metal production is
deduced from the radiation essentially emitted by massive stars. To
compare these values with direct DLA metallicity, we need to correct them
using [$\alpha$/Fe]$=0.25$ \citep{Boesgaard99}.  In our
considered metallicity range ([Fe/H]$_{mp}<-1.0$), this value is found 
approximately constant \citep{Clementini99}. 
The resulting curves were added in Fig.~\ref{gcsdlaz}.

On the one hand, the comparison of DLA and globular cluster system
metallicities is direct; both are observables.  On the other hand, a number
of assumptions have been used to estimate the metal production in the
Universe from the LBGs.  Consequently, even if these latter
curves may be used as an important check of the DLA metallicity trend, we
need to rely on the latter to settle our conclusions.

\subsection{A redshift range for the formation of the oldest globular 
clusters}
The chemical evolution of DLA systems is
below the lower limit for our metal-poor globular clusters at z
$\sim$ 4.  The conclusion suggested by these data is that the old globular
cluster formation occurred at z $<$ 4 for the adopted cosmology.

\citet{Steidel99} found that the total integrated UV luminosity
at z $\sim$ 3 is of the same order as that at z $\sim$ 4 suggesting a similar
stellar formation in the two redshift ranges. However, an 
interesting point to stress is that a small number of star-forming galaxies 
are observed at redshifts z $>$ 4 \citep[e.g.][]{Dey98,Spinrad98,Hu99}. The observed 
signal-to-noise ratio 
of these observations is low and we have only a very limited information on 
the galaxies. However, a 
preliminary conclusion is that those few galaxies may be in the very early 
stages of their formation. Nevertheless, observations of the bulk of 
high redshift ellipticals are consistent with a formation at redshifts
of the order of z $\sim 3$ to 4, while a formation at z $<$ 2 and z $>$ 5 
appears to be ruled out \citep{Treu99,Menanteau99} but see \citet{Jimenez99} 
for an alternative. To date, the detection of z $>$ 4.5 
star-forming galaxies is still anecdotal and most of the stellar formation
is observed below this redshift. The results presented in this paper 
brings an additional argument to this hypothesis. 

%%%%%%%%%%%%%%%%%%%%%%%%%%%%%%%%%%%%%%%%%%%%%%%%%%%%%%%%%%%%%%%%%%%%%%%%%
%%%%%%%%%%%%%%%%%%%%%%%%%%%%%%%%%%%%%%%%%%%%%%%%%%%%%%%%%%%%%%%%%%%%%%%%%

\section{Conclusion}

The mean metallicity of metal-poor globular clusters is
approximately constant in all galaxies and environments, with no
significant dependence of galaxy size of metallicity.  This argues
for a formation of all metal-poor globular clusters in very similar gas
fragments. Further, it suggest either a very homogeneous metallicity of the 
initial gas out of which old metal-poor globular clusters formed, and/or
very similar self-enrichment processes of the clouds.  
Self-enrichment, however, is unlikely to play an important role during the 
formation of clusters, \citep[e.g.][]{Ashman98,Nakasato00}, so that a
very homogeneous metallicity in the initial fragments is favored.

A weak correlation (to be confirmed) of the mean metallicity of metal-poor 
globular cluster systems with the host galaxy metallicity/size might
exist. This would suggest that (at least some of) the fragments in which
the metal-poor globular clusters formed were already embedded in the
dark halo of the final galaxy, rather than belonging to independent
satellites.

We found high redshift DLA systems (having high column densities of neutral 
gas, and similar metallicities to the metal-poor clusters and 
``population II'' objects) to be good 
candidates for the formation sites of metal-poor globular clusters. This
would support a picture in which, at least some, DLA systems are gas clouds 
within protogalactic halos.

\begin{acknowledgements}
We would like to thank Karl Gebhardt for his help in handling the data of
the metal poor populations and Max Pettini for helpful discussions on
DLA systems. 
Thanks also to Jane Eskdale for revising our English in the final
manuscript. Last but not least, we thank the referee Keith Ashman for
very helpful comments that improved the paper.

\end{acknowledgements}

%%%%%%%% fig 1 %%%%%%%%%%
\clearpage

%%%%%%%% fig 2 %%%%%%%%%%
\clearpage 

%%%%%%%% fig 3 %%%%%%%%%%
\clearpage 
 
%%%%%%%% fig 4 %%%%%%%%%%
\clearpage 

%%%%%%%% fig 5 %%%%%%%%%%
\clearpage 

%%%%%%%% tab 1 %%%%%%%%%%
\clearpage 
\begin{deluxetable}{lllllllll}
\tabletypesize{\scriptsize}
\tablecaption{Globular Cluster System Database.
($^*$ estimated from the published histograms) \label{tbl-1}}
\tablewidth{0pt}
\tablehead{\colhead{Galaxy} & \colhead{Type} & \colhead{(m-M)$_0$}   & \colhead{M$_V$}   &
\colhead{$(V-I)$} & \colhead{[Fe/H]$_{mp}$}  & \colhead{$\sigma_{[Fe/H]}$} 
& \colhead{N$_{GC}$} & \colhead{source}
\tablerefs{
AB93:  \citet{Ashman93}; B00:  \citet{Brodie00};
B98:  \citet{Buonanno98}; C98:  \citet{Caldwell98}; D96a:  \citet{Durell96a};
D96b:  \citet{Durell96b}; F96:  \citet{Forbes96}; F97a:  \citet{Forbes97a};
F97b:  \citet{Forbes97b}; F97c:  \citet{Forbes97c}; G96:  \citet{Geisler96};
GK99:  \citet{Gebhardt99}; H92:  \citet{Harris92}; H00:  \citet{Harris00};
KP97b:  \citet{Kissler97b}; K98:  \citet{Kundu98}; K99:
\citet{Kundu99}; K99p: \citet{Kundu99p}; M96:  \citet{Mighell96}; M98:  \citet{Montegriffo98}; N99:
\citet{Neilsen99}; O98:  \citet{Olsen98}; P99:  \citet{Puzia99}; SGHS98 :
\citet{Sarajedini98}; S95:  \citet{Secker95}; S92:  \citet{Suntzeff92}; SRN98:
\citet{Smith98}; T95:  \citet{Testa95}; W92:  \citet{Walker92}; Z95:  \citet{Zepf95}}
}
\startdata
MW      & Sbc      & 15.0 & -21.3 & 0.90    & -1.61 & 0.34 & 90 & H00 \\
LMC     & Sm       & 18.7 & -18.4 & & -1.68 & 0.32 & 12 & 
M96/T95/O98/S92/W92 \\
Sgr     & dSph     & 17.1 &       & & -1.90 & 0.10 & 2 & SRN98/M98 \\
Fornax  & dE0      & 20.6 & -13.7 & & -2.04 & 0.18 & 4 & B98 \\
NGC147  & dE5      & 24.6 & -15.1 & & -2.20 & 0.30 & 2 & M96 \\
NGC185  & dE3      & 24.6 & -15.4 & & -1.72 & 0.44 & 5 & M96 \\
NGC205  & dE5      & 24.6 & -16.5 & & -1.52 & 0.23 & 6 & M96 \\
M31     & Sb       & 24.6 & -21.8 & & -1.54 & 0.10 & 95 & AB93 \\
F8D1    & dE       & 28.0 & -14.3 & & -1.80 & 0.25 & 1 & C98 \\
M33     & Scd      & 24.5 & -19.4 & & -1.64 & 0.30 & 2 & SGHS99 \\
NGC584  & E4       & 31.7 & -21.4 & & -1.30 & 0.25 & ... &  K99p \\
NGC1023 & E/S0     & 30.0 & -21.3 & 1.02 & -1.16 & 0.25 & 65 & GK99 \\
NGC1380 & S0       & 31.1 & -19.3 & & -1.40 & 0.26 & 180 & KP97b \\
NGC1399 & E1/cD    & 31.1 & -21.6 & 0.99 & -1.26 & 0.26 & 200 & O98 \\
NGC1404 & E1       & 31.1 & -21.1 & & -1.60 & 0.15 & 372 & F97a \\
NGC1427 &    E5    & 31.3 & -20.4 & & -1.56 & 0.25 & ... & K99p \\
NGC1439 &    E1    & 31.8 & -20.4 & & -1.33 & 0.25 & ... & K99p \\
NGC2434 & E        & 30.9 & -20.1 & 0.98 & -1.36 & 0.25 & 23 & GK99 \\
NGC3115 & S0       & 30.2 & -21.3 & 0.96 & -1.36 & 0.15 & 64 & K98 \\
NGC3115DW1 & dE1.N & 30.2 & -17.7 & & -1.16 & 0.30 & 59 & D96b \\
VCC1254 & dE0.N    & 31.0 & -16.4 & & -1.45 & 0.30 & 22 & D96a \\
VCC1386 & dE3.N    & 31.0 & -16.9 & & -1.45 & 0.30 & 17 & D96a \\
NGC3311 & cD       & 33.4 & -22.3 & 0.91 & -1.50 & 0.30 & $>700$ & B00 \\
NGC3377 &    E5-6  & 30.1 & -19.9 & & -1.36 & 0.25 & ... & K99p \\
NGC3379 &    E1    & 30.1 & -20.9 & & -1.43 & 0.25 & ... & K99p \\
NGC3923 & E3       & 31.9 & -22.0 & & -0.94 & 0.15 & 138 & Z95 \\
NGC4278 &    E1LIN & 31.1 & -21.0 & & -1.46 & 0.25 & ... & K99p \\
NGC4365 & E2       & 31.4 & -21.9 & 0.95/1.04 & -1.46 & 0.25 & 160/85$^*$ & GK99/N99 \\
NGC4406 &    E3/S0 & 31.0 & -22.2 & & -1.30 & 0.25 & ... & K99p \\
NGC4458 & E0-1     & 29.9 & -18.1 & 0.83/none/none & -1.79 & 0.25 & 6$^*$ & N99/GK99/K99p \\
NGC4472 & E2       & 31.0 & -22.6 & 0.93/0.99/0.92 & -1.35 & 0.20 & 1154/282/... & G96, P99/N99/K99p \\
NGC4473 & E5       & 31.0 & -20.9 & 0.99/0.93 & -1.36 & 0.25 & 50$^*$/... & N99/K99p \\
NGC4478 & E2       & 31.4 & -20.3 & 0.99 & -1.26 & 0.30 & 53$^*$ & N99 \\
M87     & E0       & 31.0 & -22.4 & 0.96/0.98/0.95 & -1.26 & 0.25 & 296/70$^*$/382 & GK99/N99/K99 \\
NGC4486B&    cE0   & 31.0 & -17.7 & & -1.52 & 0.25 & ... & K99p \\
NGC4494 & E0       & 31.4 & -21.8 & 0.91 & -1.46 & 0.25 & 50$^*$/... & F96/K99p \\
NGC4526 & S0       & 31.0 & -21.4 & 0.88 & -1.68 & 0.25 & 28 & GK99 \\
NGC4550 & SB0      & 29.7 & -18.2 & 0.97/none/none & -1.33 & 0.25 & 17$^*$/... & N99/GK99/K99p \\
NGC4552 & E0       & 31.3 & -21.8 & 1.05/0.96 & -1.20 & 0.25 & 70$^*$/... & N99/K99p \\
NGC4594 & Sa       & 30.7 & -23.1 & & -1.20 & 0.30 & 378 & F97b \\
NGC4621 & E5       & 31.1 & -21.6 & 1.06/0.98 & -1.17 & 0.25 & 75$^*$/... & N99/K99p \\
NGC4649 & E2       & 31.0 & -22.2 & 0.99/1.02/0.95 & -1.26 & 0.25 & 120$^*$/86/... & GK99/N99/K99p \\
NGC4660 & E6       & 30.9 & -19.9 & 0.99/0.93 & -1.36 & 0.25 & 20$^*$/... & N99/K99p \\
NGC5128 & E0p      & 28.3 & -22.0 & & -1.20 & 0.15 & 42$^*$ & H92 \\
NGC5846 & E0       & 32.0 & -22.0 & 0.94/0.96 & -1.49 & 0.25 &218 & GK99/F97b \\
NGC5982 &    E3    & 32.9 & -21.7 & & -1.26 & 0.25 & ... & K99p \\
IC1459  & E3       & 31.6 & -21.9 & .../0.97 & -1.20 & 0.25 & 70$^*$/...  & F96/K99p \\
\enddata
\tablecomments{The number of globular clusters used was not recorded by 
\citet{Kundu99p} but it is typically of the order of 100 or more and never
below 20.} 
\end{deluxetable}

%%%%%%%% tab 2 %%%%%%%%%%
\clearpage 
\begin{deluxetable}{lllll}
\tabletypesize{\scriptsize}
\tablecaption{Galaxy properties for our sample. 
Mg$_2$ and $\sigma$ where taken from the
{\tt HYPERCAT} database \citep{Prugniel96,Golev98}, 
the environment density $\rho$ from \citet{Tully88}. \label{tbl-2}}
\tablewidth{0pt}
\tablehead{\colhead{Galaxy} & \colhead{$\sigma$} & \colhead{[km/s]}   
& \colhead{Mg$_2$}   & \colhead{$\rho$}
}
\startdata
Milky Way &               &                 &  0.55\\
LMC       &               &                 &  0.55\\
Sag       &    11$\pm$  4 &                 &  0.55\\
Fornax    &      6$\pm$ 3 &                 &  0.55\\
NGC147    &    22$\pm$  4 &                 &  0.55\\
NGC185    &    25$\pm$  8 & 0.095$\pm$ 0.016&  0.55\\
NGC205    &    20$\pm$  5 & 0.080$\pm$ 0.008&  0.52\\
M31       &   173$\pm$ 15 & 0.320$\pm$ 0.020&  0.52\\
F8D1      &               &                 &     \\
M33       &               &                 &  0.52\\
NGC584    &   225$\pm$ 20 & 0.292$\pm$ 0.004&  0.42\\
NGC1023   &   212$\pm$ 15 & 0.340$\pm$ 0.010&  0.57\\
NGC1380   &   240$\pm$ 15 & 0.280$\pm$ 0.008&  1.54\\
NGC1399   &   359$\pm$ 20 & 0.370$\pm$ 0.008&  1.59\\
NGC1404   &   242$\pm$ 20 & 0.340$\pm$ 0.010&  1.59\\
NGC1427   &   170$\pm$ 20 & 0.260$\pm$ 0.010&  1.59\\
NGC1439   &   159$\pm$ 20 & 0.258$\pm$ 0.021&  0.45\\
NGC2434   &   229$\pm$ 20 & 0.280$\pm$ 0.010&  0.19\\
NGC3115   &   271$\pm$ 20 & 0.330$\pm$ 0.020&  0.08\\
NGC3115DW1&    29$\pm$ 10 & 0.208$\pm$ 0.039&  0.08\\
VCC1254   &    41$\pm$ 15 &                 &  3.31\\
VCC1386   &               &                 &     \\
NGC3311   &   190$\pm$ 20 & 0.310$\pm$ 0.010&      \\
NGC3377   &   143$\pm$ 20 & 0.290$\pm$ 0.004&  0.49\\
NGC3379   &   221$\pm$ 20 & 0.337$\pm$ 0.006&  0.52\\
NGC3923   &   241$\pm$ 15 & 0.320$\pm$ 0.020&  0.40\\
NGC4278   &   251$\pm$ 20 & 0.311$\pm$ 0.010&  1.25\\
NGC4365   &   269$\pm$ 15 & 0.330$\pm$ 0.030&  2.93\\
NGC4406   &   246$\pm$ 20 & 0.334$\pm$ 0.007&  1.41\\
NGC4458   &   101$\pm$ 20 & 0.240$\pm$ 0.010&  3.21\\
NGC4472   &   303$\pm$ 20 & 0.340$\pm$ 0.010&  3.31\\
NGC4473   &   179$\pm$ 15 & 0.320$\pm$ 0.010&  2.17\\
NGC4478   &   144$\pm$ 15 & 0.260$\pm$ 0.020&  3.92\\
M87       &   339$\pm$ 30 & 0.310$\pm$ 0.030&  4.17\\
NGC4486B  &   178$\pm$ 30 & 0.304$\pm$ 0.011&  3.92\\
NGC4494   &   155$\pm$ 20 & 0.290$\pm$ 0.010&  1.04\\
NGC4526   &   260$\pm$ 20 & 0.304$\pm$ 0.008&  2.45\\
NGC4550   &    80$\pm$ 30 & 0.191$\pm$ 0.010&  2.97\\
NGC4552   &   263$\pm$ 15 & 0.350$\pm$ 0.020&  2.97\\
NGC4594   &   251$\pm$ 20 & 0.340$\pm$ 0.010&  0.32\\
NGC4621   &   237$\pm$ 15 & 0.330$\pm$ 0.010&  2.60\\
NGC4649   &   343$\pm$ 20 & 0.370$\pm$ 0.010&  3.49\\
NGC4660   &   191$\pm$ 15 & 0.320$\pm$ 0.010&  3.37\\
NGC5128   &   129$\pm$ 15 &                 &  0.20\\
NGC5846   &   250$\pm$ 20 & 0.340$\pm$ 0.010&  0.84\\
NGC5982   &   250$\pm$ 10 & 0.302$\pm$ 0.008&   \\
IC1459    &   311$\pm$ 20 & 0.340$\pm$ 0.010&  0.28\\
\enddata
\end{deluxetable}

%%%%%%%% tab 3 %%%%%%%%%%
\clearpage 
\begin{deluxetable}{cccccccccccccc}
\tabletypesize{\scriptsize}
\tablecaption{Column Density - Weighted [Zn/H] and [Fe/H].  The
[Zn/H]$_{DLA}$ are from \citet{Pettini97} while the
[Fe/H]$_{DLA}$ are compiled from \citet{Boisse98,Lu96,Pettini99,Prochaska99}.  
Columns.  \#5 and \#10 give the number of DLA metallicities
and of corresponding limits for [Fe/H] and [Zn/H] respectively. \label{tbl-3}}
\tablewidth{0pt}
\tablehead{\colhead{z} & \colhead{dz} & \colhead{T$_{min}$}   & \colhead{T$_{max}$}   &
\colhead{DLA} & \colhead{[Fe/H]$_{DLA}$}  & \colhead{$\sigma_{[Fe/H]}$} 
& \colhead{[Zn/H]$_{DLA}$} & \colhead{$\sigma_{[Zn/H]}$} & \colhead{\# DLA}
& \colhead{[Zn/Fe]} & \colhead{$\sigma_{[Zn/Fe]}$}
}
\startdata
0.45 & 0.1 &  5.1 &  5.9 &  1 (1)&-1.30&    &     &    &      &    &    \\					
1.00 & 1.0 &  5.9 &  9.7 & 10 (1)&-1.54&0.42&-0.98&0.33& 4(0) &0.56&0.53\\
1.75 & 0.5 &  9.7 & 10.5 &  5 (0)&-1.32&0.57&-0.96&0.44& 8(2) &0.36&0.72\\
2.25 & 0.5 & 10.5 & 11.0 & 13 (0)&-1.61&0.50&-1.23&0.38&12(6) &0.38&0.63\\
2.75 & 0.5 & 11.0 & 11.4 &  4 (0)&-1.29&0.64&-1.11&0.27& 7(4) &0.18&0.69\\	
3.25 & 0.5 & 11.4 & 11.6 &  3 (0)&-1.75&0.10&-1.39&    & 3(3) &    &    \\
3.75 & 0.5 & 11.6 & 11.8 &  2 (1)&-1.95&0.20&     &    &      &    &    \\
4.25 & 0.5 & 11.8 & 12.0 &  3 (1)&-2.28&0.25&     &    &      &    &    \\
2.45 & 4.1 &  5.1 & 12.0 & 47 (4)&-1.53&0.40&-1.13&0.38&34(15)&0.40&0.55\\
\enddata			
\end{deluxetable}


\begin{thebibliography}{} 
\bibitem[Anders et al.(1989)]{Anders89} Anders E., Grevesse N. 1989, Geoch. Cosmoch. Acta, 
53, 157
\bibitem[Armandroff(1989)]{Arman89} Armandroff T.E., 1989, AJ 97, 375 
\bibitem[Armandroff \& Zinn(1988)]{ArmanZinn88} Armandroff T.E. \& Zinn
R. 1988, AJ 96, 92
\bibitem[Ashman \& Bird(1993)]{Ashman93} Ashman K.M., Bird C.M. 1993 AJ 106, 2281
\bibitem[Ashman \& Zepf(1998)]{Ashman98} Ashman K.M., Zepf S.E. 1998, Globular Cluster 
Systems, Cambridge University Press
\bibitem[Ashman et al.(1994)]{Ashman94} Ashman K.M., Bird C.M., Zepf S.E. 1994 AJ 108, 2348
\bibitem[Barger et al.(1999)]{Barger99} Barger A.J., Cowie L.L., Sanders D.B. 1999, ApJ 
518, 5
\bibitem[Barmby et al.(2000)]{Barmby00} Barmby P., Huchra J.P.,
Brodie J.P., Forbes D.A., Schroder L.L. \& Grillmair C.J. 2000, AJ 119, 727
% \bibitem[Barmby \& Huchra(2000b)]{Barmby00b} Barmby, P. \& Huchra,
% J.P., 2000b ApJ 531, L29
\bibitem[Boiss\'e et al.(1998)]{Boisse98} Boiss\'e P., Le Brun V., Bergeron J., Deharveng 
J.-M.  1998, A\&A 333, 841
\bibitem[Boesgaard et al.(1999)]{Boesgaard99} Boesgaard A.M., King J.R., Deliyannis C.P., Vogt 
S.S. 1999, A.J. 117, 492
\bibitem[Brodie \& Huchra(1991)]{Brodie91} Brodie J.P., Huchra J.P. 1991 ApJ 379,157
\bibitem[Brodie et al.(2000)]{Brodie00} Brodie J.P., Larsen S.S., Kissler-Patig M.
2000, ApJL submitted
\bibitem[Buonanno et al.(1998)]{Buonanno98} Buonanno R., Corsi C. E., Zinn R. et al. 1998, ApJ 501, L33
\bibitem[Caldwell et al.(1998)]{Caldwell98} Caldwell N., Armandroff T.E., da Costa G.S., 
Seitzer P. 1998, AJ 115,535
\bibitem[Chaboyer et al.(1998)]{Chaboyer98} Chaboyer B., Demarque P., Kernan P.J., Krauss 
L.M. 1998, ApJ 494, 96 
\bibitem[Clementini et al.(1999)]{Clementini99} Clementini G., Gratton R.G., Carretta E., 
Sneden C. 1999, MNRAS 302, 22
\bibitem[C\^ot\'e et al.(1998)]{Cote98} C\^ot\'e P., Marzke R.O., West M.J. 1998, ApJ 501, 
554
\bibitem[Cohen et al.(1998)]{Cohen98} Cohen J.G., Blakeslee J.P., Ryshov A.  1998, ApJ 
496, 808
\bibitem[Dey et al.(1998)]{Dey98} Dey A., Spinrad H., Stern D.,  Graham J.R., Chaffee F.H. 1998, ApJ 498, L93
\bibitem[Dressler et al.(1987)]{Dressler87} Dressler A., Lynden-Bell D., Burstein D., et
al. 1987, ApJ 313, 42
\bibitem[Durell et al.(1996a)]{Durell96a} Durell P.R., Harris W.E., Geisler D., Pudritz 
R.E. 1996a, AJ 112, 972
\bibitem[Durell et al.(1996b)]{Durell96b} Durell P.R., McLaughlin D.E., Harris W.E. , Hanes 
D.H. 1996b, ApJ 463., 543
\bibitem[Feltzing \& Gilmore(2000)]{Feltzing00} Feltzing S., Gilmore G. 2000, A\&A 355, 949
\bibitem[Forbes et al.(1996)]{Forbes96} Forbes D.A., Franx M., Illingworth G.D., Carollo 
C.M. 1996, ApJ 467,126
\bibitem[Forbes et al.(1997a)]{Forbes97a} Forbes D.A., Brodie J.P., Grillmair C.J.
1997a, AJ 113, 1652
\bibitem[Forbes et al.(1997b)]{Forbes97b} Forbes D.A., Grillmair C.J., Smith R.C.
1997b, AJ 113, 1648
\bibitem[Forbes et al.(1997c)]{Forbes97c} Forbes D.A., Brodie J.P., Huchra J.P. 
1997c, AJ 113, 887
\bibitem[Gebhardt \& Kissler-Patig(1999)]{Gebhardt99} Gebhardt K., Kissler-Patig M. 1999 AJ 118, 1526
\bibitem[Geisler et al.(1996)]{Geisler96} Geisler D., Lee M.G., Kim E. 1996, AJ 111, 1529
\bibitem[Golev \& Prugniel(1998)]{Golev98} Golev V., Prugniel P. 1998, A\&AS 132, 255
\bibitem[Haehnelt et al.(1996)]{Haehnelt96} Haehnelt  M. G., Steinmetz M., Rauch, M. 1996 ApJ 
465, L95
\bibitem[Harris(2000)]{Harris00} Harris W.E. 2000, in "Lectures for the 1998 SAAS-FEE 
Advanced Course on Star Clusters" (Swiss Society for Astrophysics and 
Astronomy), in press, (http://physun.mcmaster.ca/~harris/WEHarris.html)
\bibitem[Harris et al.(1992)]{Harris92} Harris G.L.H., Geisler D., Harris H.C., Hesser J.E.  1992, AJ 104, 613
\bibitem[Hilker et al.(1999)]{Hilker99} Hilker, M., Infante, L., and Richtler, T. 1999, A\&A Suppl., 138, 55
\bibitem[Hu et al.(1999)]{Hu99} Hu E.M., McMahon R,G,m Cowie L.L. 1999, ApJ 522, L9
\bibitem[Ibata et al.(1994)]{Ibata94} Ibata R, Gilmore G.F., Irwin M.J. 1994, Nat. 370,194
\bibitem[Jimenez \& Padoan(1998)]{Jimenez98} Jimenez R., Padoan P. 1998, ApJ 498, 704
\bibitem[Jimenez et al.(1999)]{Jimenez99} Jimenez R., Friaca A.C.S., Dunlop J.S. et al. 1999, MNRAS 305, L16
\bibitem[Katz et al.(1996)]{Katz96} Katz N., Weinberg D.H., Hernquist L., Miralda-Escude 
J. 1996, ApJ 457, L57
\bibitem[Kennicutt(1989)]{Kennicutt89} Kennicutt R.C. 1989, ApJ 344, 685
\bibitem[Kinman(1959)]{Kinman59} Kinman T.D. 1959, MNRAS 119, 559
\bibitem[Kissler-Patig et al.(1997a)]{Kissler97a} Kissler-Patig M., Richlter T., Storm J., Della 
Valle M. 1997a, A\&A 327, 503
\bibitem[Kissler-Patig et al.(1997b)]{Kissler97b} Kissler-Patig M., Kohle S., Hilker M. et al. 1997b, A\&A 319, 83
\bibitem[Kissler-Patig et al.(1998a)]{Kissler98a} Kissler-Patig M., Forbes D.A., Minniti D., 
1998a, MNRAS 298, 1123
\bibitem[Kissler-Patig et al.(1998b)]{Kissler98b}  Kissler-Patig M., Brodie J.P., Schroder L.L, et 
al. 1998b, ApJ 115, 105
\bibitem[Kobulnicky \& Koo(2000)]{Kobulnicky00} Kobulnicky H.A., Koo D. 2000, astro-ph/0008242
\bibitem[Kundu (1999)]{Kundu99p} Kundu A., 1999, PhD Univ.~Maryland
\bibitem[Kundu \& Whitmore(1998)]{Kundu98} Kundu A., Whitmore B.C. 1998, AJ 116, 2841
\bibitem[Kundu et al.(1999)]{Kundu99} Kundu A., Whitmore B.C., Sparks W.B. et al. 1999, ApJ 513, 733
\bibitem[Lee et al.(1998)]{Lee98} Lee M.G., Kim E., Geisler D. 1998, ApJ 115, 947

\bibitem[Lowenthal et al.(1997)]{Lowenthal97} Lowenthal J.D., Koo D.C., Guzman R. et al. 1997, 
ApJ 481, 673
\bibitem[Lu et al.(1996)]{Lu96} Lu L., Sargent W.L.W., Barlow T.A., Churchill C.W., 
Vogt  S.S. 1996, ApJS 107, 475
\bibitem[Menanteau et al.(1999)]{Menanteau99} Menanteau F., Ellis R.S., Abraham R.G., Barger A.J., Cowie L.L. 1999, 
MNRAS 309, 208
\bibitem[Mighell et al.(1996)]{Mighell96} Mighell K.J., Rich R.M., Shara M., Fall S.M. 1996, AJ 111, 2314
\bibitem[Minniti(1995)]{Minniti95} Minniti D. 1995, AJ 109, 1663
\bibitem[Montegriffo et al.(1998)]{Montegriffo98} Montegriffo P., Bellazzini M., Ferraro F.R. et al. 1998, MNRAS 294, 315
\bibitem[Nakasato et al.(2000)]{Nakasato00} Nakasato N., Mori M., Nomoto K. 2000, ApJ 535, 776
astro-ph/0001333
\bibitem[Neilsen \& Tsvetanov(1999)]{Neilsen99} Neilsen E.H., Tsvetanov Z.I. 1999, ApJ 515, L13
\bibitem[Olsen et al.(1998)]{Olsen98} Olsen K.A.G., Hodge P.W., Mateo M. et al. 1998, MNRAS 300, 665
\bibitem[Ortolani et al.(1995)]{Ortolani95} Ortolani S., Renzini A., Gilmozzi R. et al. 1995, Nat. 377, 701
\bibitem[Olszewski et al.(1996)]{Olszewski96} Olszewski E.W., Suntzeff N.B., Mateo D. 1996, 
ARAA 34, 511
\bibitem[Parmentier et al.(2000)]{Parmentier00} Parmentier G., Jehin E., Magain P., Noels A., Thoul A.A. 2000, astro-ph0009477
\bibitem[Pei \& Fall(1995)]{Pei95} Pei C., Fall S.M. 1995, ApJ 454, 69
\bibitem[Pettini et al.(1997)]{Pettini97} Pettini M., Smith L.J., King D.L., Hunstead R.W.  
1997, ApJ 486, 665
\bibitem[Pettini et al.(1999)]{Pettini99} Pettini M., Ellison S.L., Steidel C.C., Bowen D.V. 
1999, ApJ 510, 576
\bibitem[Prochaska \& Wolfe(1999)]{Prochaska99} Prochaska J.X., Wolfe A.M. 1999 ApJS 121,3691
\bibitem[Prochaska \& Wolfe(2000)]{Prochaska00} Prochaska J.X., Wolfe A.M. 2000, ApJ 533, L5
\bibitem[Prugniel \& Simien(1996)]{Prugniel96} Prugniel P., Simien F. 1996, A\&A 309, 749
\bibitem[Puzia et al.(1999)]{Puzia99} Puzia T.H., Kissler-Patig M., Brodie J.P., 
Huchra J.P. 1999, AJ 118, 2734
\bibitem[Renzini(1999)]{Renzini99} Renzini A. 1999, astro-ph/9902108
\bibitem[Richtler (1994)]{Richtler94} Richtler T., 1994, Reviews in
Modern Astronomy, Vol.~8, ed.G.Klarer, P.163 
\bibitem[Rosenberg et al.(1999)]{Rosenberg99} Rosenberg A., Saviane I., Piotto G., Aparicio A 1999, AJ 118, 2306
\bibitem[Sarajedini et al.(1998)]{Sarajedini98} Sarajedini A., Geisler D., Harding P., Schommer R. 1998, ApJ 508, L37
\bibitem[Searle \& Zinn (1978)]{Searle78} Searle L., \& Zinn R., 1978, ApJ 225, 357
\bibitem[Secker et al.(1995)]{Secker95} Secker J., Geisler D., McLaughlin D.E., Harris W.E.
1995, AJ 109,1019
\bibitem[Smith et al.(1998)]{Smith98} Smith E.O., Rich R.M., Neill J.D. 1998, AJ 115, 2369
\bibitem[Spinrad et al.(1998)]{Spinrad98} Spinrad H., Stern D., Bunker A. et al. 1998, AJ 116, 2617
\bibitem[Steidel et al.(1996)]{Steidel96} Steidel C.C., Giavalisco M., Adelberger K.L., 
Dickinson M.  1996, ApJ 462, L17
\bibitem[Steidel et al.(1999)]{Steidel99} Steidel C.C., Adelberger K.L., Giavalisco M., 
Dickinson M., Pettini M. 1999, ApJ 519,1
\bibitem[Suntzeff et al.(1992)]{Suntzeff92} Suntzeff N.B., Schommer R.A., Olszewski E.W.,
Walker A.R. 1992, AJ 104, 1743
\bibitem[Testa et al.(1995)]{Testa95} Testa V., Ferraro F.R., Brocato E., Castellani V. 
1995, MNRAS 275, 454
\bibitem[Treu \& Stiavelli(1999)]{Treu99} Treu T., Stiavelli M. 1999, AJ 524, L27
\bibitem[Tully(1988)]{Tully88} Tully R.B. 1998, Nearby Galaxies Catalogue,
Cambridge University Press
\bibitem[vandenBergh(1975)]{vandenBergh75} van den Bergh S. 1975, ARA\&A 13,217
\bibitem[Walker(1992)]{Walker92} Walker A.R. 1992, AJ 104, 1395
\bibitem[Wolfe et al.(1995)]{Wolfe95} Wolfe A.M., Lanzetta K.M., Foltz C.B., Chaffee F.H. 
1995, ApJ 454, 698
\bibitem[Zepf \& Ashman(1993)]{Zepf93} Zepf  S.E., Ashman K.M. 1993, MNRAS 264, 611
\bibitem[Zepf et al.(1995)]{Zepf95} Zepf  S.E., Ashman K.M., Geisler D. 1995, ApJ, 443, 570
\bibitem[Zinn(1985)]{Zinn85} Zinn R. 1985, ApJ, 293, 424
\bibitem[Zinn(1993)]{Zinn93} Zinn R. 1993, "Stellar Populations", eds.  C.A.  
Norman, A.  Renzini and M.  Tosi, Cambridge Univ. Press, p.73 
\end{thebibliography}
\end{document}